\documentclass{jkas}

\def\year{2022} 
\def\month{December} 
\def\volume{55} 
\def\issue{6} 
\def\beginpage{207} 
\def\endpage{213} 
\def\received{October 16, 2022} 
\def\accepted{November 22, 2022} 
\date{Received \received; accepted \accepted}

\usepackage{flushend} 

\usepackage{epstopdf}

\usepackage[colorlinks,
            pdftex,
            breaklinks,
            linkcolor=blue,
            urlcolor=blue,
            anchorcolor=blue,
            menucolor=blue,
            citecolor=blue]{hyperref}
\usepackage{amsmath, graphicx, natbib}
\usepackage{array,booktabs,multirow, makecell}

\newcolumntype{x}[1]{>{\raggedright\arraybackslash}m{#1}}
\newcolumntype{y}[1]{>{\centering\arraybackslash}m{#1}}
\newcolumntype{z}[1]{>{\raggedleft\arraybackslash}m{#1}}
\title{
Renovation of Seoul Radio Astronomy Observatory \\ and Its First Millimeter VLBI Observations}

\author[1,2]{\href{https://orcid.org/0000-0002-9598-0018}{\mbox{Naeun Shin}}}
\author[1,3]{\href{https://orcid.org/0000-0003-0198-3136}{\mbox{Yong-Sun Park}}}
\author[2,4]{\mbox{Do-Young Byun}}
\author[1]{\mbox{Jinguk Seo}}
\author[1]{\mbox{Dongkok Kim}}
\author[1]{\mbox{Cheulhong Min}}
\author[2]{\mbox{Hyunwoo Kang}}
\author[5]{\mbox{Keiichi Asada}}
\author[5]{\mbox{Wen-Ping Lo}}
\author[1,3]{\href{https://orcid.org/0000-0003-0465-1559}{\mbox{Sascha Trippe}}}

\affil[1]{Department of Physics and Astronomy, Seoul National University, 1, Gwanak-ro, Gwanak-gu, Seoul~08826, \mbox{Republic of Korea}; \email{neshin@kasi.re.kr}}
\affil[2]{Korea Astronomy and Space science Institute, 776, Daedeok-daero, Yuseong-gu, Daejeon~34055, \mbox{Republic of Korea}}
\affil[3]{SNU Astronomy Research Center, Seoul National University, 1, Gwanak-ro, Gwanak-gu, Seoul~08826, \mbox{Republic of Korea}}
\affil[4]{University of Science and Technology, 217, Gajeong-ro, Yuseong-gu, Daejeon~34113, \mbox{Republic of Korea}}
\affil[5]{Academia Sinica Institute of Astronomy and Astrophysics, Taipei~10617, Taiwan}

\def\corrauthor{
Y.-S. Park
}

\def\runningauthor{
Shin et~al.
}

\def\runningtitle{
Renovation of SRAO and Its First VLBI Observation
}

\def\keywords{
instrumentation: interferometers --- techniques : high angular resolution
}

\def\abstracttext{
The Seoul Radio Astronomy Observatory (SRAO) operates a 6.1-meter radio telescope on the Gwanak campus of Seoul National University. We present the efforts to reform SRAO to a Very Long Baseline Interferometry (VLBI) station, motivated by recent achievements by millimeter interferometer networks such as Event Horizon Telescope, East Asia VLBI Network, and Korean VLBI Network (KVN). For this goal, we installed a receiver that had been used in the Combined Array for Research in Millimeter-wave Astronomy and a digital backend, including an H-maser clock. The existing hardware and software were also revised, which had been dedicated only to single-dish operations. After several years of preparations and test observations in 1 and 3-millimeter bands, a fringe was successfully detected toward 3C~84 in 86~GHz in June 2022 for a baseline between SRAO and KVN Ulsan station separated by 300~km. Thanks to the dual frequency operation of the receiver, the VLBI observations will soon be extended to the 1~mm band and verify the frequency phase referencing technique between 1 and 3-millimeter bands.
}

\begin{document}
\jkashead 


\section{Introduction\label{sec:intro}}
Since its inauguration in 2001 in a 3~mm band, the 6-meter telescope of the Seoul Radio Astronomy Observatory (SRAO) has been actively used in research and education \citep{koo03}.
The antenna design is identical to the one used for the Berkeley-Illinois-Maryland Association array \citep{hudson98}.
The surface is adjusted by Holographic surface measurements to an accuracy of around 30 $\mu$m, enabling observations up to 300~GHz \citep{lee03}.
In 2013, SRAO developed a 1~mm band receiver with a side band separation feature and a receiver temperature of 50~K in a single sideband (SSB) \citep{lee13}.
After several years of operations of the receiver, it had suffered from troubles in cooling systems, suspending regular operation.

Meanwhile, the Event Horizon Telescope (EHT) presented the first image of the black hole in M87 at 230~GHz, highlighting the crucial role of the Very Long Baseline Interferometry (VLBI) \citep{eht19}.
One of the major science goals of the next generation EHT is the imaging and monitoring of the jet launching region in M87 and other active galactic nuclei (AGNs), aiming at a more detailed understanding of the physics in the ultimate vicinity of supermassive black holes.
For this aim, the EHT collaboration considers the upgrades of the VLBI network that include a wider observing bandwidth, higher angular resolution using higher frequencies and possibly longer baselines when extended to space, and a more dense UV-coverage through the addition of numerous smaller telescopes.
Scientists in east Asia have also established East Asia VLBI Network operating in the 1~mm band, called EAVN-hi.
The Korean VLBI Network (KVN) tested the performance of its Yonsei antenna with a 1~mm receiver.
The \mbox{SRAO-KVN} Yonsei pair is separated only by about 10~km, and it could provide a short baseline for the recovery of extended features.
The fourth telescope of KVN is under construction, designed to operate up to 230~GHz with an aperture efficiency of around 30\%.

For the possible participation in the next generation EHT and EAVN-hi and the collaborations with the extended KVN, SRAO retrofitted the observing system that had been set to single-dish mode operations.
We describe the new receiver and the revisions of the observing system in Section~\ref{sec:sec2}.
The preparation of VLBI backends is introduced in Section~\ref{sec:sec3}.
Section~\ref{sec:sec4} presents the test observation in single-dish mode and the first fringe detection in the VLBI experiments.
The results are summarised and discussed in Section~\ref{sec:sec5}.

\section{Retrofit of the Receiver System\label{sec:sec2}}
\subsection{Import of a CARMA Receiver\label{sec:sec21}}

Since the trouble with the 4~K cryogenic system mentioned above was severe, we had to consider purchasing a new one. We finally decided to recycle a receiver that had once operated in the Combined Array for Research in Millimeter-wave Astronomy (CARMA).
The main reasons for adopting the CARMA receiver are that it is paired with an economical 15~K cooling system, and its performance is still good.
The only drawback may be that the CARMA receiver works in double sidebands (DSB) mode.
It features dual-band operation in the 1~mm and the 3~mm bands, receiver temperatures of 60--70~K (DSB), and compactness \citep{hull15}.

\begin{figure}[!t]
\centering
\includegraphics[width=\linewidth]{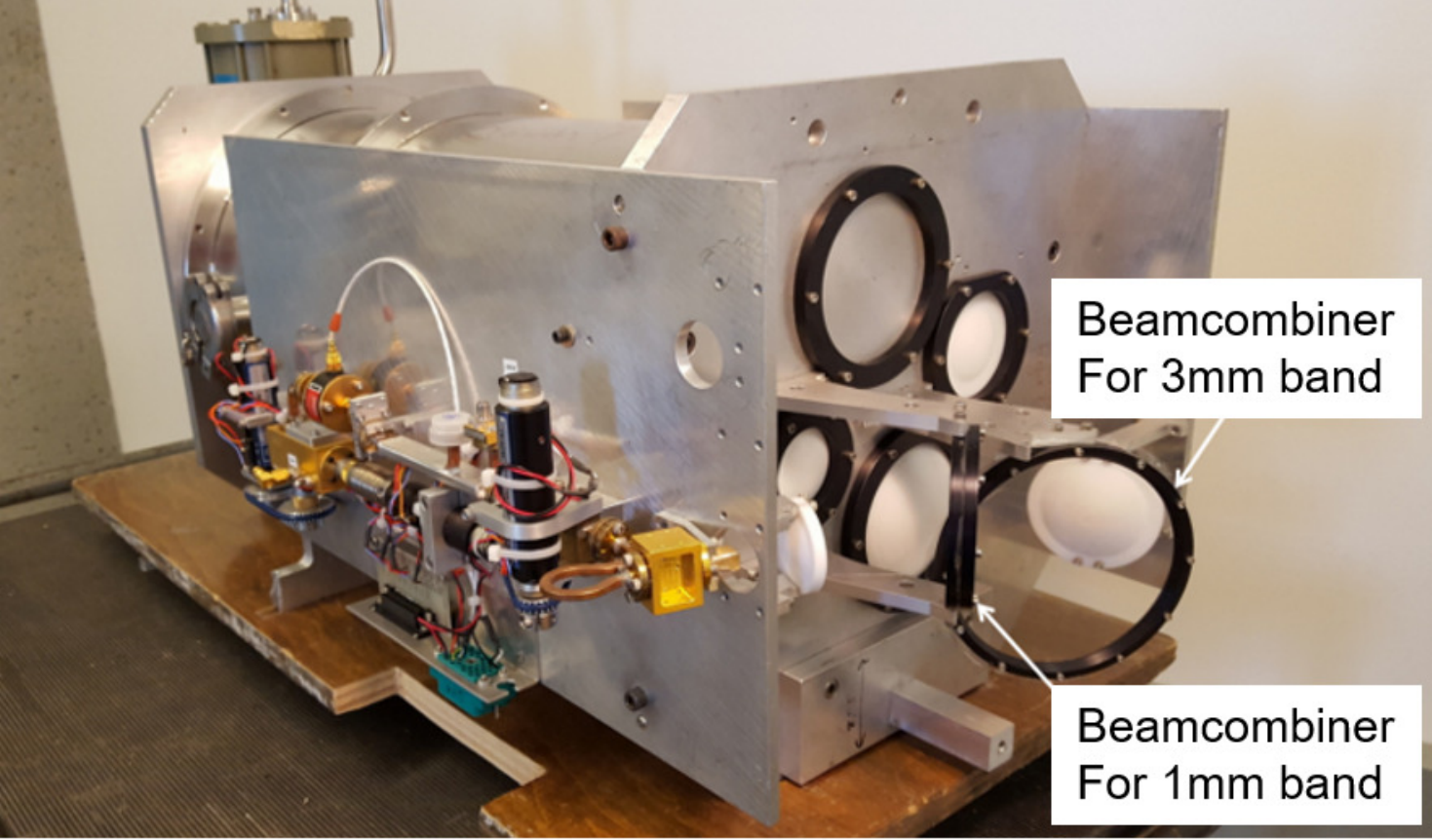}
\caption{The view of the CARMA receiver.
Teflon lenses and beam combiners for the two bands are shown together. 
The LO plates are mounted on both sides of the dewar. (Photo by Richard Plambeck)\label{srao02}}
\vspace{0.5mm}
\end{figure}

\begin{figure}[!t]
\centering
\includegraphics[width=\linewidth]{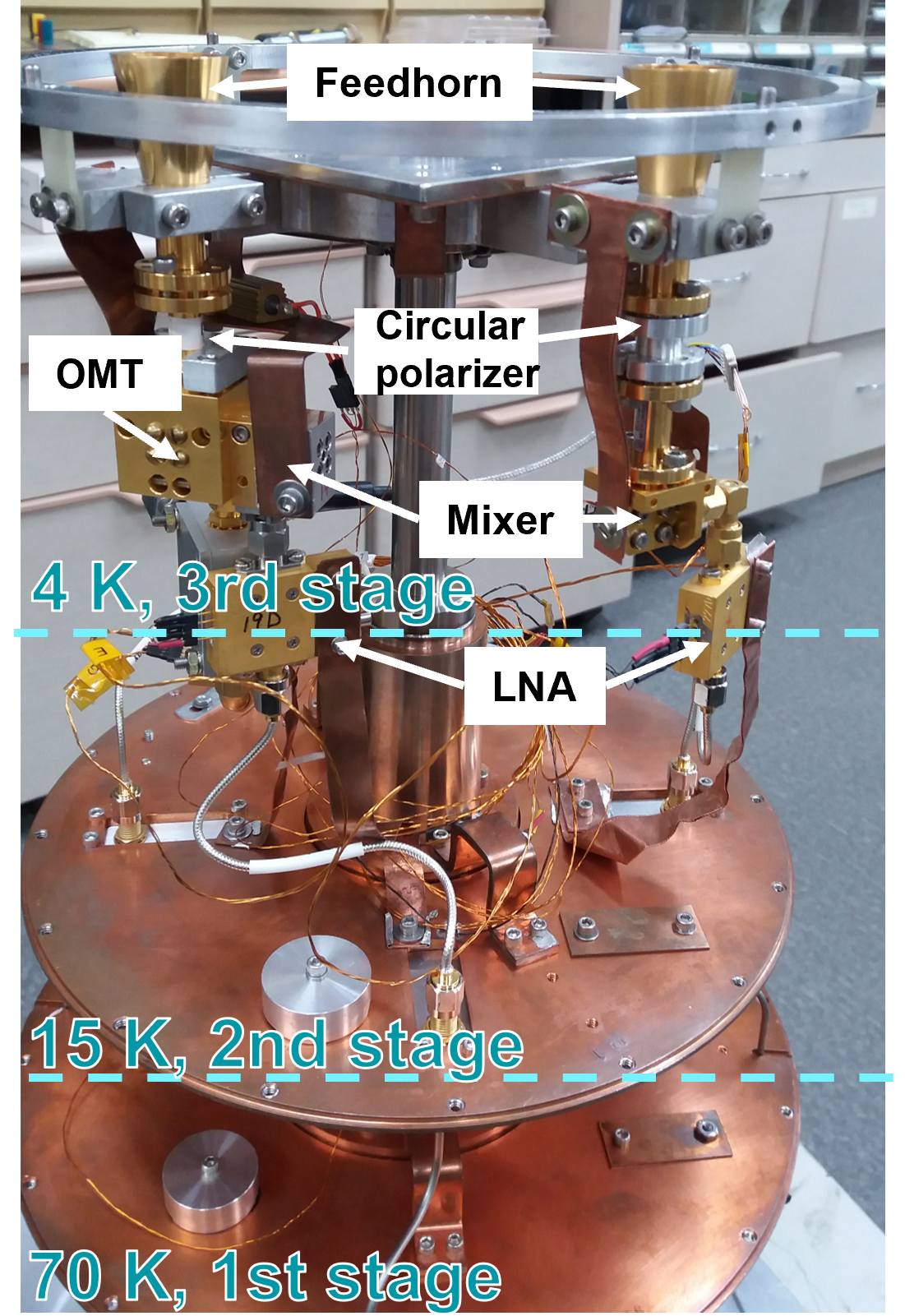}
\caption{The cryogenically cooled dewar contains the 1~mm band receiver parts on the left side and the 3~mm band parts on the right side. 
The copper plates at the bottom are the first and second cooling stages, and the 3rd is the square plate at the top of the pillar in the center.
Feed horns, circular polarizers, an OMT, and mixers are cooled below 4~K, while LNAs are cooled to $\sim$15~K.\label{srao01}}
\vspace{5mm}
\end{figure}

The CARMA receiver is displayed in Figure~\ref{srao02}.
Millimeter waves from the antenna pass through two Teflon lenses without complicated beam-guiding optics and go into the dewar. The interior of the dewar is divided mainly into two parts---left for the 1~mm band and right for the 3~mm band, as shown in Figure~\ref{srao01}.
The part of the 1~mm band consists of a feed horn, a circular polarizer, an orthomode transducer (OMT), superconductor-insulator-superconductor (SIS) mixers, and low-noise amplifiers (LNA) \citep{hull15}.
This configuration allows dual circular polarization observations.
The 3~mm band receiving system is similar but does not have an OMT, allowing only the right-handed circular polarization (RCP).
Table \ref{tab:rxtable1} summarizes the receiver parameters in the two bands.
Local oscillator (LO) plates at both sides of the dewar generate the LO signals for the two bands, as shown in Figure~\ref{srao02}.
The mylar beam combiners reflect the LO signals in front of the dewar.
Then the LO signals combined with waves from celestial sources propagate into the dewar.

\begin{figure}[!t]
\centering
\includegraphics[width=\linewidth]{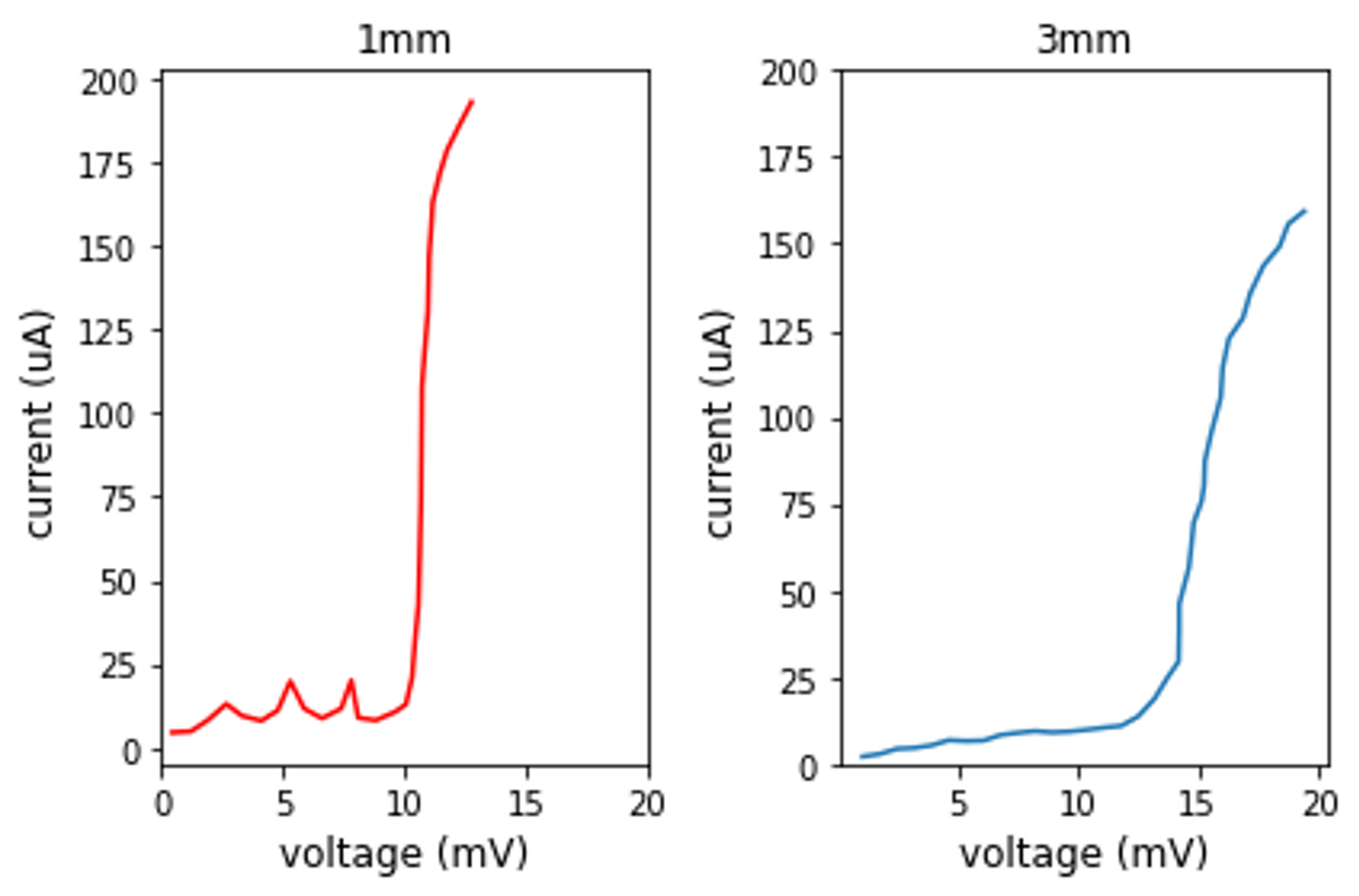}
\caption{The unpumped I-V curves of the 1~mm and the 3~mm mixers were measured in the laboratory. They exhibit a nice non-linearity.
\label{srao08}}
\vspace{5mm}
\end{figure}

After we took the receiver to the laboratory, we tested its components one by one and measured receiver temperatures representing its overall performance.
Figure~\ref{srao08} displays the I-V curve of the 1~mm and the 3~mm mixers without LO power.
One can see a steep current increase at bias voltages of around 10 mV.
The measured receiver temperatures are similar to the ones in \citet{hull15}.

\begin{table}[!t]
\caption{Receiver parameters \label{tab:rxtable1}}
\centering\small
\begin{tabular}{y{0.4\linewidth} *{2}{y{0.3\linewidth}}}
\toprule
& 1~mm & 3~mm \\
\midrule
RF frequency & 215--270~GHz & 84--115~GHz\\
IF bandwidth & \multicolumn{2}{c}{0--1~GHz}\\
Polarization & LCP/RCP & RCP\\
Receiver temperature & \multicolumn{2}{c}{60--70~K (DSB)}\\
System temperature & 400~K (DSB) & 150~K (DSB)\\
\bottomrule
\end{tabular}
\vspace{2mm}
\end{table}

\begin{figure}[!t]
\centering
\includegraphics[width=\linewidth]{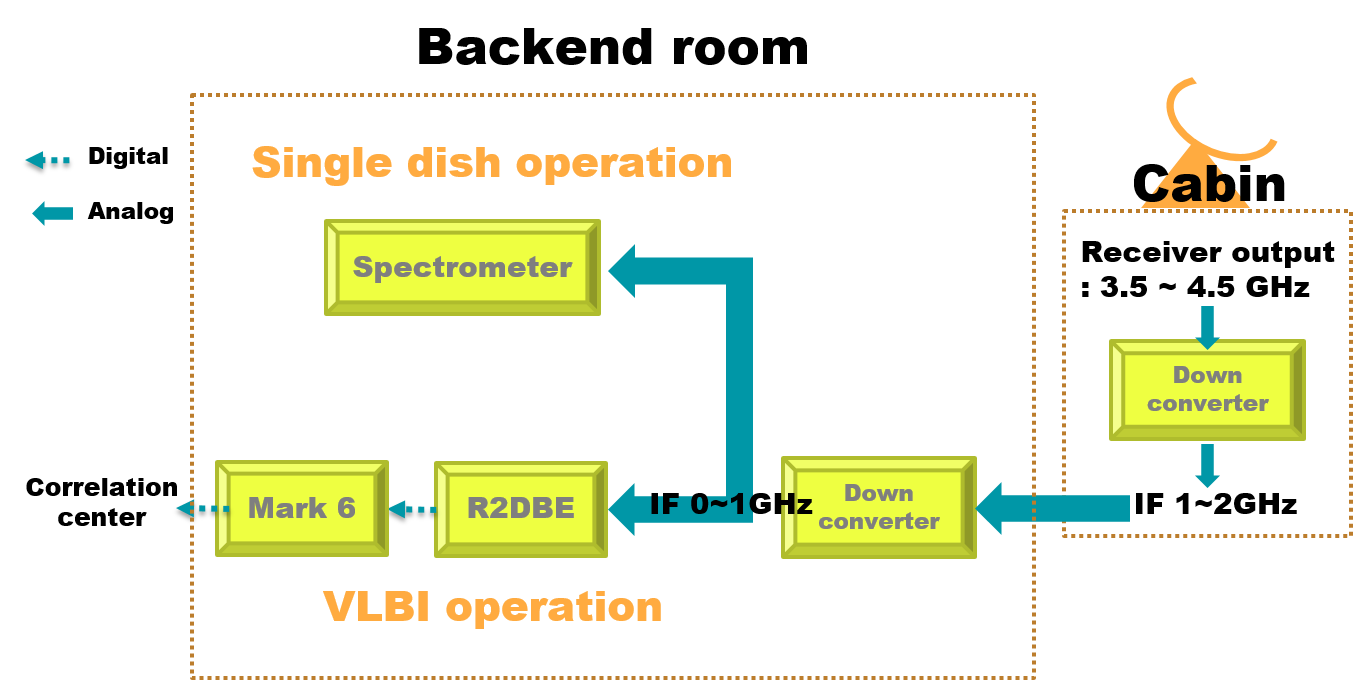}
\caption{The system diagram including the conversion of IF frequencies. The VLBI instruments and the spectrometer for single-dish observation are located in the backend room. 
\label{sig}}
\vspace{2mm}
\end{figure}

\subsection{Cryogenic System\label{sec:sec22}}

The cryogenic system consists of a cold head and a compressor.
Within the dewar is a CTI1020 cryogenic cold head that consists 
of two cooling stages, where the second stage cools down to 15~K.
It was modified to have an additional third stage by the University of California, Berkeley, which is shown in Figure~\ref{srao01}. 
The cooling power of the third stage is only 50~mW, but enough to cool down mixers and feed horns to around 4~K.
The 15~K cryogenic system is not so expensive but has gained the 4~K stage after modification, which is one of the reasons that we adopt the CARMA receiver.

The third stage can be further cooled down by lowering the pumping frequency. 
The cold head typically operates at 72~rpm, which seems the most efficient cycle frequency for the heat transfer of cryogenic regenerator \citep{ogawa91}. 
However, it is empirically found that if the pumping period is increased by a factor of two, then the temperature of the third stage further decreases below 3~K, probably because the cooling He gas spends more time inside the third stage \citep{plambeck93}. 
We usually set the pumping speed to 30~rpm using a frequency inverter, SV-is7, made by a local company, LS Electric Co.
The typical temperatures during the regular operation are 47.4, 12.2, and 2.8~K, respectively, from the first to third stages.
As for the compressor, we use the model M600 made by Trillion. 
Though a water-cooled compressor has a higher cooling capacity with a smaller volume, we adopt the air-cooled one since the air-cooled compressor requires fewer maintenance efforts.

\subsection{Signal Chain\label{sec:sec23}}
The signal's intermediate frequency (IF) bandwidth after the LNAs is rather wide but is narrowed down to 3.5--4.5~GHz through bandpass filters.
The IF signal is then down-converted to 1--2~GHz in the IF processing box in the cabin. It goes to the observatory building and is converted to 0--1~GHz before being fed into the spectrometers and VLBI backends. The spectrometer covers 1~GHz bandwidth with $2^{14}$ channels resulting in a spectral resolution of 61~kHz. 
The signal flows for single-dish and VLBI operation modes are summarized in Figure~\ref{sig}.

\subsection{Alignment of the Receiver\label{sec:sec24}}

\begin{figure}[!t]
\centering
\includegraphics[width=\linewidth, height = 71mm]{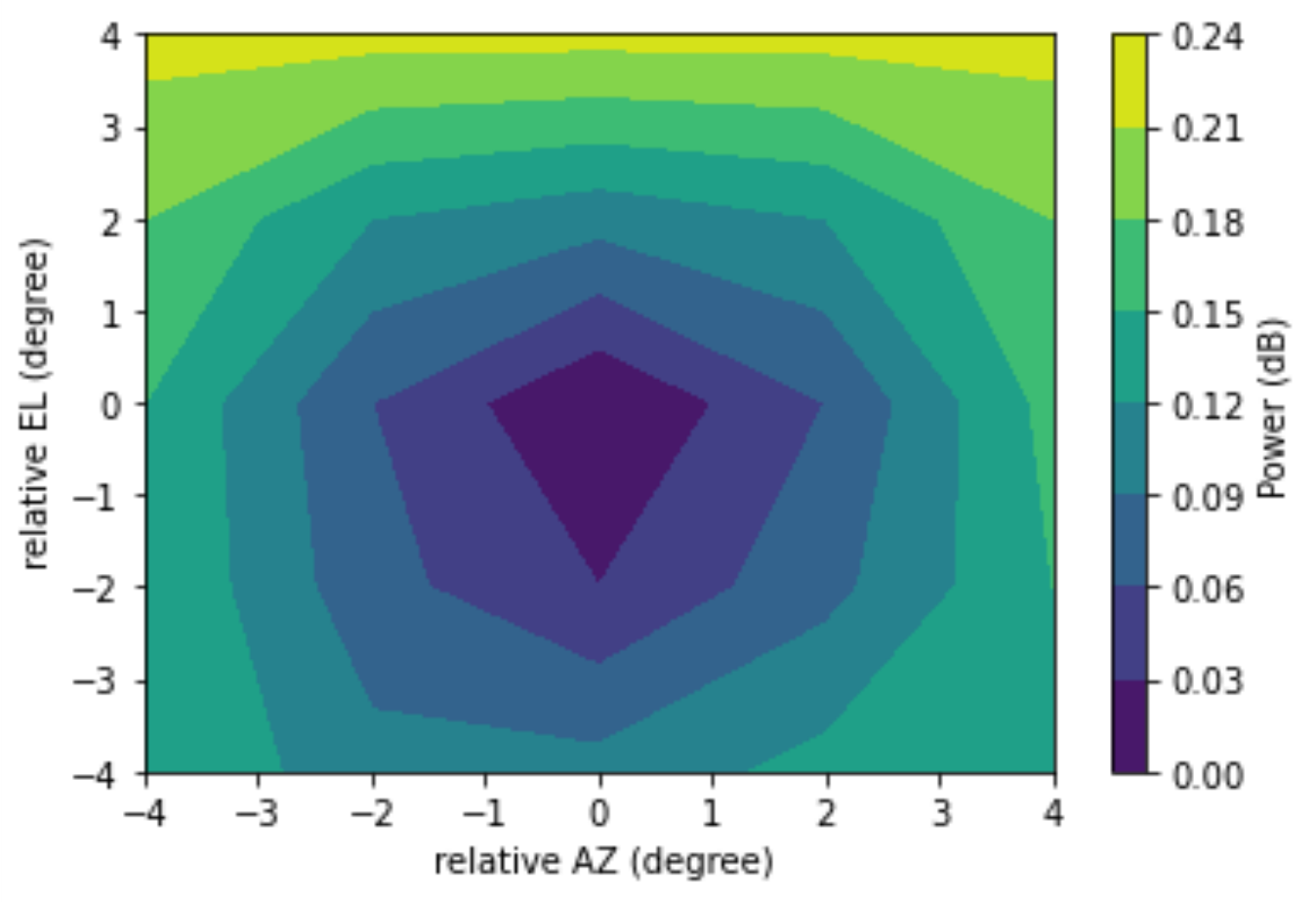}
\caption{The relative output power of the receiver in dB for various locations of the 80~K absorber. 
The coordinate origin of the map refers to the center of the subreflector.
The boresight of the feed horn is slightly shifted to the low elevation. 
\label{srao04}}
\vspace{2mm}
\end{figure}

Since the configuration changed near the secondary focus, we need to check whether the direction of the maximum gain of the feed horn points to the center of the subreflector. 
First, we point the telescope 
towards a mountain, which is seen at low elevations, i.e., the 300~K background. 
Then by moving an absorber immersed in liquid nitrogen at 80~K in front of the subreflector and by measuring the output power of the receiver, we can find the direction of the maximum response of the horn. 
Figure~\ref{srao04} indicates that the feed horn looks at the subreflector downward by $1^{\circ}$, while it is well aligned in the E-W direction.
We corrected this misalignment by adjusting the heights of the receiver plate in four corners.

The change of the receiver position also affects the pointing offsets, which must be corrected.
We established the pointing model by carrying out five-point observations for the standard stars in the Hipparcos catalog \citep{byun02}, using the optical telescope attached to the side of the antenna dish \citep{koo03}.
The {\it rms} pointing errors are around $15^{\prime\prime}$ in both directions right after the model fitting. 
However, systematic offsets begin to appear, probably due to the differential solar heating of the antenna structure, which may affect the 1~mm band observation because of its smaller beam size.

\begin{figure*}[!t]
\centering
\includegraphics[width=0.5\linewidth]{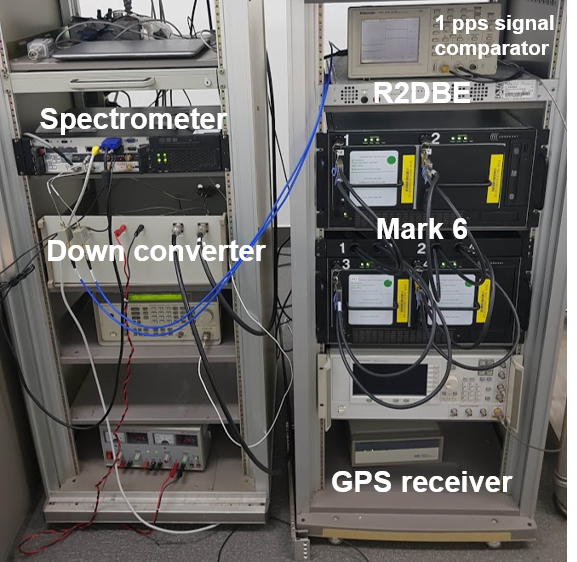}
\caption{The VLBI backends, R2DBE, Mark 6, and GPS receiver are located in the backend room of SRAO.
A spectrometer and the second IF processing box are shown together on the left.
\label{srao05}}
\vspace{5mm}
\end{figure*}

\begin{figure}[!t]
\centering
\includegraphics[width=\linewidth]{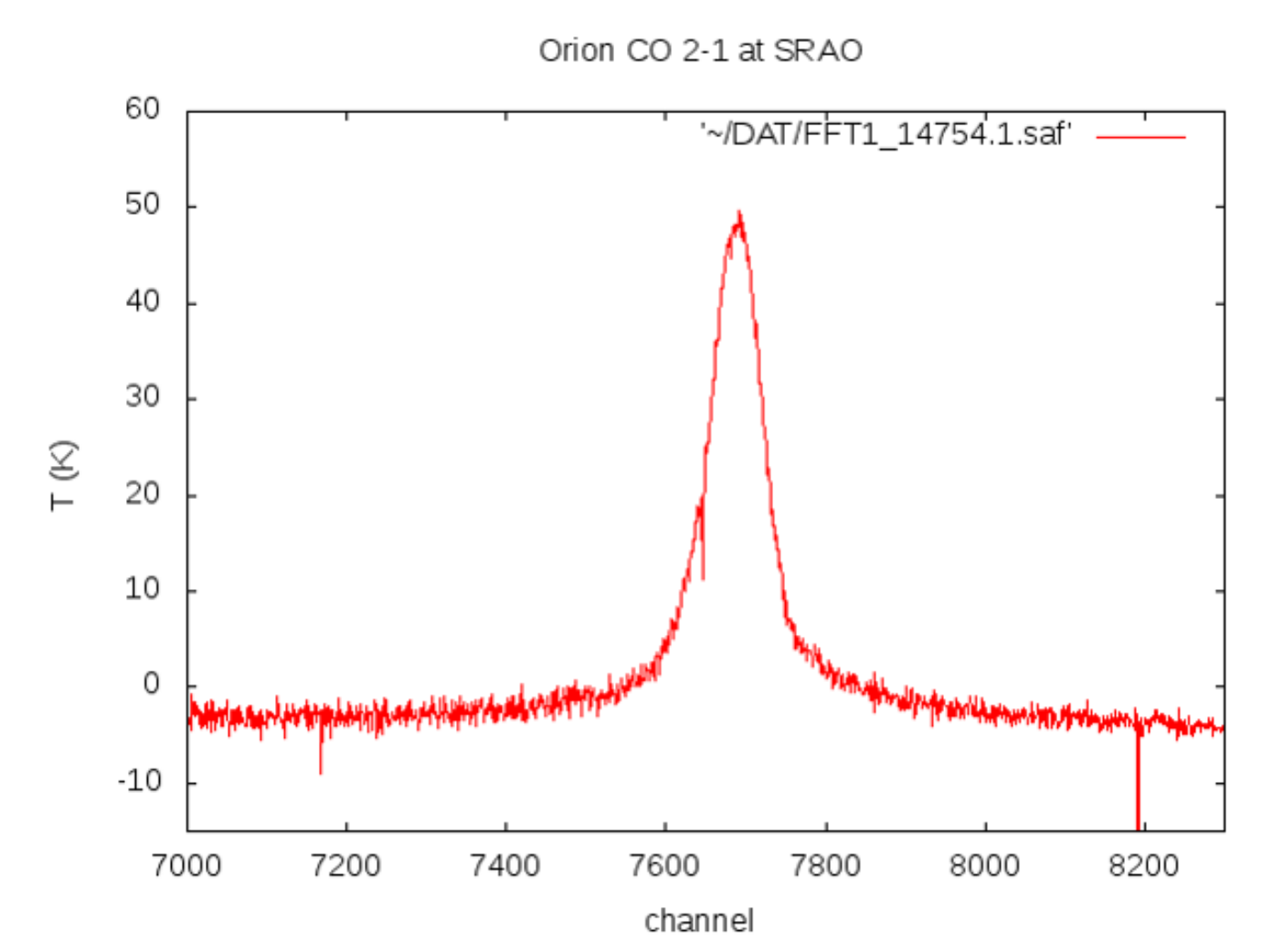}
\caption{
The first spectrum of CO $J=2-1$ at 230.538~GHz toward the Orion KL was obtained by the SRAO with the CARMA receiver. 
The temperature scale and the velocity of the spectrum are not calibrated accurately. 
A velocity resolution is 0.079~km\,s$^{-1}$ per channel.
\label{srao09}}
\vspace{2mm}
\end{figure}

\section{Installation of VLBI Equipment\label{sec:sec3}}

\begin{table*}[!t]
\caption{3~mm VLBI test observation parameters}
\label{obstable1}
\centering\small
\begin{tabular}{*{3}{y{0.33\linewidth}}}
\toprule
& SRAO & KVN \\
\midrule
Equipment & R2DBE / Mark 6 & OCTAD / Mark 6\\
Bandwidth & 1024~MHz & 2048~MHz\\
Polarization & RCP & LCP/RCP \\
System temperature & 150~K (DSB) & 280~K (SSB) \\
\bottomrule
\end{tabular}
\vspace{2mm}
\end{table*}

A digital sampler, recorders, and an H-maser clock are installed to carry out VLBI experiments in SRAO, as shown in Figure~\ref{srao05}.
As for the digital sampler, we borrowed a ROACH~2 digital backend (R2DBE) from Academia Sinica Institute of Astronomy and Astrophysics (ASIAA) \citep{vertatschitsch15}.
Its maximum data rate is 16~Gbps, from 4~Gbps samples per second in four levels for two data streams.
We also borrowed a Mark~6 recorder from ASIAA and four disk packs from the Korea Astronomy and Space science Institute (KASI) with a total capacity of 256~TBytes.
The H-maser clock, provided by KASI, distributes a reference frequency of 10~MHz to all the frequency synthesizers in the receiver system and the recorder.
An additional component, the GPS receiver, compares its one pulse per second (PPS) signal and that of the H-maser clock for synchronization with other stations.

Several frequency synthesizers in the receiving system were of low quality and independent of each other in the past since it was not so critical for single-dish observations.
For the VLBI experiment, we bought a few high-quality frequency synthesizers, such as Keysight E8257D, to reduce the phase noises of the system.
They replaced old synthesizers and are bound to the 10~MHz reference from the H-maser clock.

\section{Test Observations\label{sec:sec4}}
\subsection{Single-Dish Observations\label{sec:sec41}}
In February 2019, SRAO detected the first light of the CO $J=2-1$ line at 230.538~GHz
toward Orion KL with the CARMA receiver (Figure~\ref{srao09}). 
The best system temperature for a 1~mm band is measured as 400~K (DSB). 
It is found that, contrary to expectations, the system temperature rises in the winter.
The reduction of cooling capacity due to the hardening of oil in the compressor may cause this problem.
We expect lower system temperatures by keeping the compressor warm in the winter.

The 3~mm band receiver was operated in the spring for recent two years. 
The lowest system temperature in the 3~mm band is 150~K (DSB) at 86~GHz. 
We have made single-dish observations of several bright spectral line sources, such as Orion KL and TX-Cam, to inspect the pointing accuracy and verify the overall system performance before the VLBI observation.

\subsection{VLBI Test Observations\label{sec:sec42}}
\subsubsection{Test Observations in the 1~mm Band\label{sec:sec421}}
As soon as we got the first light at 230~GHz in 2019, we conducted international VLBI observations.
During UT 11:00 to 18:00 on March 18th and 19th, 2019, the Greenland Telescope (GLT), built by ASIAA in Taiwan, and Solar Planetary Atmosphere Research Telescope, operated by Osaka prefecture university in Japan, joined the campaign.
Targets were three bright AGNs (M87, NGC~6251, Mrk~501) and four bright calibrators (3C~371, 1928+738, 3C~345, 1633+382). 
The spectral line sources (NGC~7027, IRC+10216, DR21) are also observed for autocorrelation. 
No fringes were detected for baselines that include SRAO.
The suspected reason is that one LO accidentally missed the 10~MHz reference signal from the H-maser clock.

The second test observation was run during UT from 08:00 to 15:00 on February 1st and 5th, 2020, collaborating with GLT and James Clerk Maxwell Telescope.
Two bright AGNs (OJ~287, 3C~84) were observed, and data was transferred to the correlation center at the Shanghai Astronomical Observatory via the internet. The typical data transfer rate was around 500~Mbps.
Unfortunatly, the fringes were not detected in this session too. 
We concluded that the main reason for the failure is the substantial system temperature, probably because of the reason mentioned in Section~\ref{sec:sec41} and cloudy weather during the observation.

\subsubsection{VLBI Test Observations with KVN in the 3~mm Band\label{sec:sec422}}
\label{3mmobs}

Since the scheduling is not easy in the international VLBI observations, and thus the chance of observations is limited, we cooperated with the domestic VLBI system, the KVN.
Since the KVN does not have the 1~mm receivers, the VLBI test observation was made in the 3~mm band.
From UT 05:00 June 3rd, 2022, the bright source, 3C~84 and Orion KL, were observed alternately in three scans each, with 10~minutes of exposure per scan. 
3C~84, a bright AGN, is selected as the main target to find a fringe.
Orion KL is observed to check the frequency offset and the pointing accuracy.
In SRAO, a data stream of 1024~MHz bandwidth from 86 to 87~GHz is Nyquist-sampled and recorded in 4~Gbps.
On the other hand, KVN stations recorded signals of 2048~MHz bandwidth from 85 to 87~GHz for two polarization in 16~Gbps using an OCTAD sampler \citep{oh17} and Mark 6 recorder. The system setup is summarized in Table \ref{obstable1}.
Because of the instrumental problems, only the KVN Ulsan station recorded the data among the three KVN stations.
The recorded data in the Mark 6 was transferred to the Daejeon correlation center located at KASI through the internet. The DiFX Software correlator performed the correlation in 65~K channels.

\begin{figure}[!t]
\centering
\includegraphics[width=\linewidth, trim={0mm -3mm 0mm 0mm, clip}]{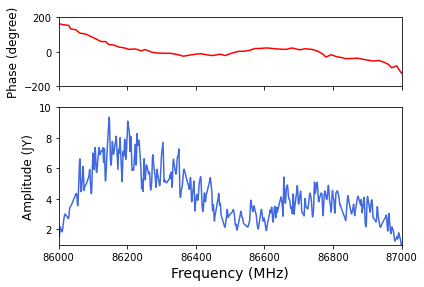}
\caption{The visibility phase (top) and the cross power spectrum (bottom) of 3C~84 for the SRAO to KVN Ulsan baseline. 
\label{srao011}}
\vspace{2mm}
\end{figure}

Visibility data is analyzed with the NRAO Astronomical Image Processing System (AIPS). The fringe fitting solutions are found using the task FRING of AIPS with a solution interval of 30~seconds. 
The upper panel of Figure~\ref{srao011} shows the clear and consistent phase as a function of frequency after the fringe fitting. The cross-power spectrum between SRAO and KVN Ulsan is displayed together.

\begin{figure}[!t]
\centering
\includegraphics[width=\linewidth, trim={0mm -3mm 0mm 0mm, clip}]{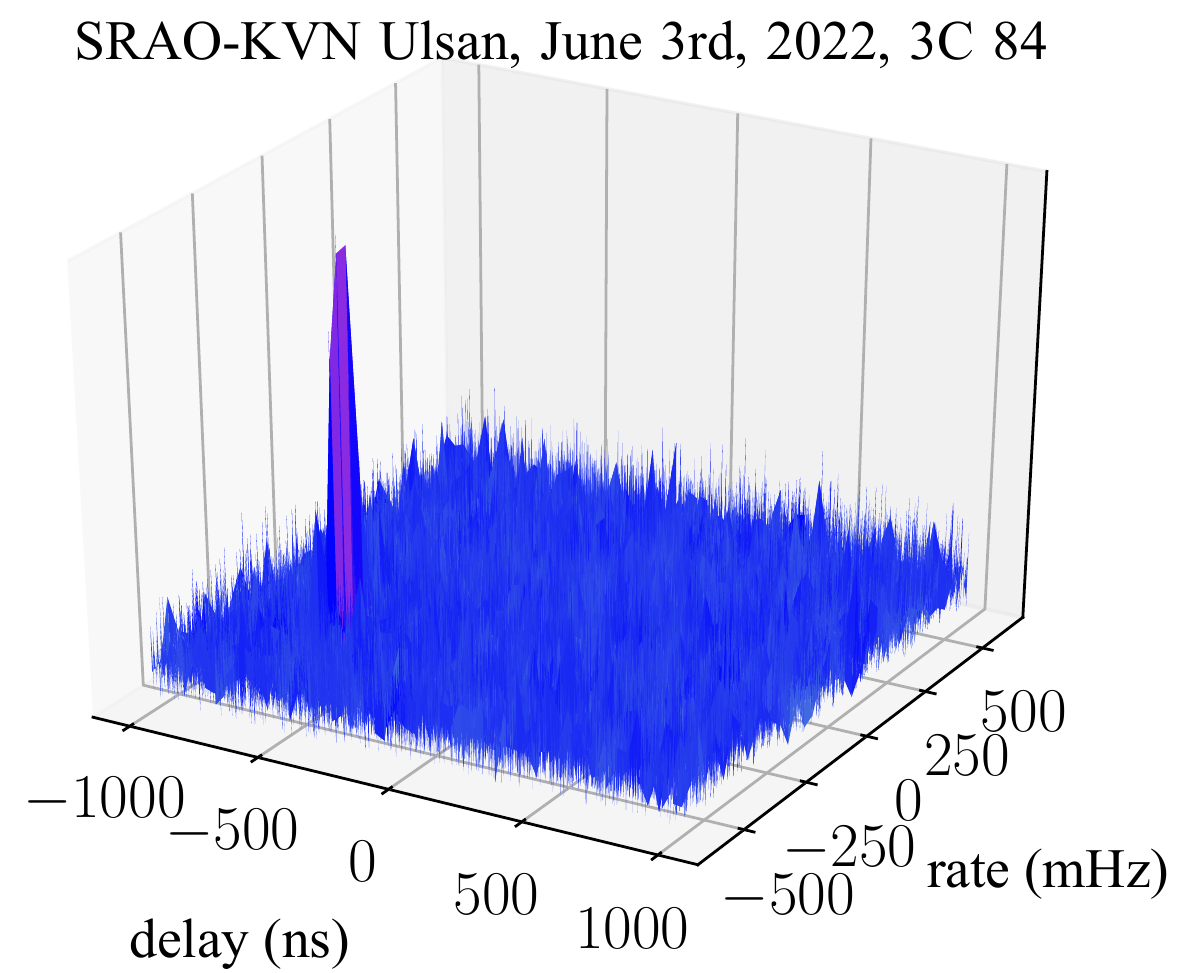}
\caption{The visibility amplitude averaged over the central 640~MHz bandwidth as a function of the delay and the delay rate.
The delay and the delay rate at the peak are about $-580$~ns and $-250$~mHz, respectively. 
\label{srao013}}
\vspace{2mm}
\end{figure}

Figure~\ref{srao013} presents a fringe solution in a delay and delay rate plane. The visibility amplitude is averaged over the central 640~MHz of the bandwidth, where phases remain constant.
We can see a strong peak for a specific delay and delay rate.
The obtained delay rate of 250~mHz between SRAO and KVN Ulsan is mainly due to the frequency offset of about 200~mHz of the H-maser clock at KVN Ulsan station. 

The observed SNRs are 60 on average for a solution interval of 30~seconds. 
We can estimate the expected SNR using the equation, 
\begin{equation*}
\label{eq:snr}
{\rm SNR} = 0.88\,F\,\sqrt{\frac{2\,\Delta\nu\,\tau}{{\rm SEFD}_1 \times {\rm SEFD}_2}},
\end{equation*}
where $F$ is a source flux density, $\tau$ the integration time, and $\Delta\nu$ the observing bandwidth. The ${\rm SEFD}_i$ is the system equivalent flux density of a station $i$.
We set $\Delta \nu=640$~MHz and $\tau=30$~seconds.
The F is assumed as 16~Jy based on the single-dish mode observation of KVN toward 3C~84 in May 2022. 
The SEFD of the KVN is 3200~Jy. The SEFD of the SRAO is not measured, but it can be accurately inferred as $4.1 \times 10^{4}$~Jy from the comparison of the antenna temperatures of the SiO $v=1$, $J=2-1$ transition toward Orion KL obtained by both KVN in single-dish mode and SRAO on the same day. The resulting SNR is 120, much larger than the observed one.

The factor of two difference may be originated from the two reasons.
The first one is a longer averaging time to derive visibility data from raw data streams from the two antennas. We set one~second for it as usual, but the delay rate of $-250$~mHz makes the phase rotate by $90 ^{\circ}$ for one~second, which results in a factor of $2 / \pi$ degradation of visibility amplitudes. 
The other one is related to the source size. According to the 86~GHz observation of 3C~84 with KVN, the core and the jet components are separated in north-south direction by about 3~mas \citep{wajima20}. The SRAO-KVN Ulsan baseline length of 300~km results in the minimum fringe spacing of 2.5~mas, and thus the source might be partially resolved out. 
Figure~\ref{srao011} shows a flux density of $\sim$5~Jy, weaker than that of single-dish observation indeed. 
The solution interval comparable to the coherence time of the atmosphere may also affect the SNR. However, it is found that data reduction with the solution interval of 10~seconds does not improve the SNR. 

\section{Discussion and Conclusions\label{sec:sec5}}

The SRAO retrofitted the receiver and cooling systems, enabling observations in the 1~mm and the 3~mm bands with reasonable noise temperature.
We also installed instruments such as a digital backend and high-speed recorder to reform SRAO to a VLBI station.
After many years of single-dish observation and international VLBI campaigns, we detected fringes at 86~GHz between KVN Ulsan and SRAO in June 2022 for the first time. 
Since the LO chain of the 1~mm band is identical to that of the 3~mm band, except for the frequency tripler, it is expected to find fringes in the 1~mm band in the near future.

To make routine VLBI observations possible, we need to improve our system further: A new receiver is under development, adopting sideband separation mixers and a new cryostat. 
It will widen the IF bandwidth from 1~GHz to 2~GHz and have a lower noise temperature than the CARMA receiver.
Moreover, SRAO will be connected to Korea Research Environment Open Network (KREONET), and the data transfer rate will be over 10~Gbps.
In addition, with the help of \mbox{KREONET}, a 10~MHz reference signal from KVN Yonsei may be transmitted to SRAO via dark fibers, which will replace the H-maser clock operated over its life span. 

VLBI observations at millimeter wavelengths are considerably affected by rapid atmospheric phase variations. 
This infection can be minimized by applying the solutions of the phase variations taken at the lower frequencies to the higher target frequencies \citep{rioja11, rioja14, algaba15, park18}.
SRAO can implement the frequency phase referencing with minimal effort, since the 1~mm and the 3~mm receiving components are in one dewar, and LOs share a common frequency reference. 
The only thing to do is to install a frequency-selective surface and a reflection mirror in front of the dewar.

In summary, a series of test observations and planned future works guarantee that SRAO can be a member of the international~mm VLBI network.

\acknowledgments
The authors gratefully acknowledge the contribution of Jongho Park for providing a code of the 3D plot of fringes and a kind explanation.
We are grateful to the staff of the KVN who helped to operate the array and to correlate the data. The KVN and a high-performance computing cluster are facilities operated by the KASI. The KVN observations and correlations are supported through the high-speed network connections among the KVN sites provided by the KREONET, which is managed and operated by the KISTI.
This work was supported partially by National Research Foundation of Korea grant funded by the Korean government (MEST) (No.~2019R1A6A1A10073437 and 2022R1F1A1075115) and partially by KASI under the R\&D program (Project No. 2022-1-860-03) supervised by the Ministry of Science and ICT.



\begin{thebibliography}{}


\bibitem[Algaba et~al.(2015)]{algaba15} 
Algaba, J.-C., Zhao, G.-Y., Lee, S.-S., et~al. 2015, Interferometric monitoring of Gamma–Ray bright Active Galactic Nuclei II: Frequency Phase Transfer, JKAS, 48, 237

\bibitem[Byun \& Yun(2002)]{byun02} 
Byun, D., \& Yun, Y.-Z. 2002, Development of control softwares for the SRAO 6-meter telescope based on PCs running Linux, Exp. Astron., 14, 183

\bibitem[The Event Horizon Telescope collaboration et al.(2019)]{eht19} 
The Event Horizon Telescope Collaboration, Akiyama, K., Alberdi, A., Alef, W., et~al. 2019, First M87 Event Horizon Telescope Results. I. The Shadow of the Supermassive Black Hole, ApJL, 875, L1

\bibitem[Hudson(1998)]{hudson98} 
Hudson, J. 1998, Ray-tracing the BIMA reflectors, BIMA Memoranda Series, No.~64

\bibitem[Hull \& Plambeck(2015)]{hull15} 
Hull, C., \& Plambeck, R. 2015, The 1.3~mm full-stokes polarization system at CARMA, J. Astron. Instrum., 4, 1550005

\bibitem[Koo et~al.(2003)]{koo03}
Koo, B., Park, Y.-S., Hong, S., et~al. 2003, Performance of the SRAO 6-meter radio telescope, JKAS, 36, 43

\bibitem[Lee et~al.(2003)]{lee03} 
Lee, S.-S., Byun, D.-Y., Park, Y.-S., et~al. 2003, Surface Adjustment of the SRAO 6-M Antenna Based on Near-Field Radio Holography at 86~GHz, International journal of infrared and millimeter waves, 24, 1687

\bibitem[Lee et~al.(2013)]{lee13}
Lee, J.-W., Kim, C.-H., Kang, H., et~al. 2013, Development of 230~GHz radio receiver system for SRAO, JKAS, 46, 225

\bibitem[Oh et~al.(2017)]{oh17} 
Oh, S.-J., Yeom, J.-H., Roh, D.-K., et~al. 2017, A Study on the Test Results of 32~Gbps Observing System for Wideband VLBI Observation, Journal of the institute of signal processing and systems, 18, 13

\bibitem[Ogawa et~al.(1991)]{ogawa91} 
Ogawa, M., Li, R., \& Hashimoto, T. 1991, Thermal conductivities of magnetic intermetallic compounds for cryogenic regenerator, Cryogenics, 31, 405

\bibitem[Park et~al.(2018)]{park18}
Park, J., Kam, M., Trippe, S., et~al. 2018, Revealing the Nature of Blazar Radio Cores through Multifrequency Polarization Observations with the Korean VLBI Network, ApJ, 860, 112

\bibitem[Plambeck et~al.(1993)]{plambeck93}
Plambeck, R., Thatte, N., \& Sykes, P. 1992, A 4K Gifford-Mcmahon refrigerator for radio astronomy, 7th International cryocooler conference proceeding, 2, 401

\bibitem[Rioja et~al.(2011)]{rioja11}
Rioja, M.~J., Dodson, R., Malarecki, J., et~al. 2011, Exploration of Source Frequency Phase Referencing Techniques for Astrometry and Observations of Weak Sources with High Frequency Space Very Long Baseline Interferometry, AJ, 142, 157

\bibitem[Rioja et~al.(2014)]{rioja14}
Rioja, M.~J., Dodson, R., Jung, T., et~al. 2014, Verification of the Astrometric Performance of the Korean VLBI Network, Using Comparative SFPR Studies with the VLBA at 14/7~mm, AJ, 148, 84

\bibitem[Vertatschitsch et~al.(2015)]{vertatschitsch15}
Vertatschitsch, L., Primiani, R., Young, A., et~al. 2015, R2DBE: A Wideband Digital Backend for the Event Horizon Telescope, PASP, 127, 1226

\bibitem[Wajima et~al.(2020)]{wajima20}
Wajima, K., Kino, M., \& Kawakatu, N. 2020, Constraints on the Circumnuclear Disk through Free-Free Absorption in the Nucleus of 3C~84 with KaVA and KVN at 43 and 86~GHz, ApJ, 895, 35



\end{thebibliography}
\end{document}